\RequirePackage{fix-cm}
\documentclass[smallcondensed]{svjour3}     % onecolumn (ditto)
\smartqed  % flush right qed marks, e.g. at end of proof
\usepackage{mathptmx}
\usepackage{helvet}
\usepackage{courier}
\usepackage{amsmath,amssymb}
\usepackage{amsfonts}
\usepackage{braket}
\usepackage{amsbsy}
\usepackage{bm}
\usepackage{type1cm}  
\usepackage{dcolumn}       
\usepackage{index}         % allows index generation
\usepackage{graphicx} 
\usepackage{tikz}
\usepackage{multicol}        % used for the two-column index
\usepackage[bottom]{footmisc}% places footnotes at page bottom
\usepackage{epstopdf}
\usepackage{subfigure}
\usepackage{mathtools}
\usepackage{color}
\usepackage{bigints}
\mathtoolsset{centercolon}
\usepackage{bm}
\usepackage{stmaryrd}

\newcommand{\eV}{e^{(V)}}
\newcommand{\eE}{{e}^{(E)}}

\newcommand{\rstar}{\rho^*}
\newcommand{\jump}[1]{\llbracket #1 \rrbracket}

\usepackage{mathptmx}   
\begin{document}

% \title{Acceleration Waves and the K-Condition in Nonlinear Hyperbolic Models of Viscoelastic Solids and Non-Newtonian Fluids}
\title{Acceleration Waves and the K-Condition in Viscoelastic Solids and Non-Newtonian Fluids}

\titlerunning{Acceleration waves in viscoelasticity and non-Newtonian fluids}

      % if too long for running head

\author{Tommaso Ruggeri   
}


\institute{Tommaso Ruggeri \at
            Department of Mathematics University of Bologna \& \\
                Accademia dei Lincei, Roma \\
                 \email{tommaso.ruggeri@unibo.it}
}

\date{}
\dedication{With great esteem and fraternal affection, this paper is dedicated to Marco Sammartino on the occasion of his 60th birthday.}

\maketitle
\begin{abstract}
The K-condition introduced by Shizuta and Kawashima provides a sufficient criterion for the global existence of smooth solutions to dissipative hyperbolic systems. For genuinely nonlinear characteristic fields, a weaker K-condition becomes necessary, although not sufficient. In this paper, we analyze this weaker K-condition through the study of acceleration waves propagating in an equilibrium state.
We investigate two classes of hyperbolic models: one describing viscoelasticity with linear dissipation, and the other non-Newtonian fluids asymptotically converging to a power-law behavior. For viscoelastic models, the weaker K-condition is always satisfied and acceleration waves remain bounded. For non-Newtonian fluids, the validity of the condition depends on the power-law index $m$: it holds for Newtonian fluids ($m=1$), is violated for shear-thinning fluids ($m<1$), and 
leads to an instantaneous regularization of acceleration waves  for shear-thickening fluids ($m>1$).
\end{abstract}

% \begin{abstract}
% The K-condition introduced by Shizuta and Kawashima provides a sufficient criterion for the global existence of smooth solutions to dissipative hyperbolic systems. For genuinely nonlinear characteristic fields, a weaker K-condition becomes necessary, though not sufficient. In this paper, we analyze this weaker K-condition through the study of acceleration waves propagating in an equilibrium state.
% We investigate two classes of hyperbolic models: one describing viscoelasticity with linear dissipation, and the other non-Newtonian fluids asymptotically converging to a power-law behavior. For viscoelastic models, the weaker K-condition is always satisfied and acceleration waves remain bounded. For non-Newtonian fluids, the validity of the condition depends on the power-law index $m$: it holds for Newtonian fluids ($m=1$), is violated for shear-thinning fluids ($m<1$), and leads to finite-time regularization of acceleration waves for shear-thickening fluids ($m>1$).
% \keywords{Acceleration waves \and Hyperbolic systems with relaxation \and Viscoelasticity \and Non-Newtonian fluids}
% \end{abstract}

\section{Introduction}
In the theory of hyperbolic and hyperbolic--parabolic systems of conservation laws, the existence of a strictly convex entropy function is a fundamental requirement for well-posedness. In particular, symmetric systems of conservation laws admit a unique local-in-time smooth solution for smooth initial data \cite{FeL,Kawa,fisher}.

In general, however, global continuation of smooth solutions cannot be ensured, even for arbitrarily small smooth initial data. Finite-time formation of singularities, shocks, or blow-up may occur \cite{maida,dafermos}.

In many physically relevant models, the interplay between hyperbolicity and relaxation can prevent the appearance of singularities. The source term introduces a dissipative mechanism competing with hyperbolicity. If dissipation dominates, the system is said to be \emph{dissipative}, and global smooth solutions converging to equilibrium are expected. If dissipation and hyperbolicity are comparable, only part of the perturbation diffuses and the system is referred to as \emph{composite} \cite{zeng}.

Different criteria have been proposed to distinguish dissipative from composite systems. One approach, analogous to that used for hyperbolic--parabolic systems, was introduced by Shizuta and Kawashima \cite{Kawa2}. This condition is usually referred to as the \emph{K-condition} or \emph{genuine coupling} condition.

For one-dimensional dissipative systems satisfying the K-condition, global existence of smooth solutions for small initial data was established by Hanouzet and Natalini \cite{nat}, Yong \cite{wen}, and Bianchini, Hanouzet, and Natalini \cite{bnat}. Moreover, Ruggeri and Serre \cite{RugSerre} proved the stability of constant equilibrium states. However, the K-condition is only a sufficient criterion; global existence may still hold if it is violated, as shown by Zeng \cite{zeng}.

Lou and Ruggeri \cite{Palermo} showed that a weaker K-condition provides a necessary, though not sufficient, criterion for global existence. While the classical K-condition requires that all right eigenvectors of the hyperbolic part do not belong to the kernel of the gradient of the production term, the weaker K-condition only concerns eigenvectors associated with genuinely nonlinear characteristic fields. It was verified that these hypotheses hold in both classical \cite{Ruget} and relativistic \cite{RugETR,RUGCHO} Rational Extended Thermodynamics for monatomic gases, as well as in multi-temperature gas mixtures \cite{SR}.

This weaker K-condition admits a natural interpretation in terms of discontinuity waves, or, in continuum mechanics language, acceleration waves. If the weaker K-condition is violated, any nonzero initial acceleration amplitude leads to a finite critical time at which the amplitude becomes unbounded.

The purpose of the present paper is to further investigate this weaker K-condition by studying acceleration waves in recently proposed nonlinear symmetric hyperbolic models for viscoelasticity and non-Newtonian fluids. In particular, we show that for viscoelastic models the weaker K-condition is always satisfied, whereas for non-Newtonian fluids asymptotically converging to a power-law model, its validity strongly depends on the power-law index $m$. This results in distinct behaviors of acceleration waves: bounded decay for viscoelastic solids and shear-thickening fluids, and finite-time blow-up for shear-thinning and, practically, Newtonian fluids.

\section{Balance laws, K-condition and global existence}

Physical laws in continuum theories are expressed in the form of balance laws. For simplicity, we restrict ourselves to one spatial dimension:
\begin{equation}  \label{conservation}
\partial_t \mathbf{u} + \partial_x \mathbf{G}(\mathbf{u}) = \mathbf{f}(\mathbf{u}),
\end{equation}
where $\mathbf{u}(x,t)$, $\mathbf{G}(\mathbf{u})$, and $\mathbf{f}(\mathbf{u})$ are $\mathbb{R}^N$-valued vectors, and $\partial_t$ and $\partial_x$ denote partial derivatives with respect to time $t$ and the spatial variable $x$, respectively.

The production term $\mathbf{f}(\mathbf{u})$ represents dissipation. However, not all of its components are necessarily nonzero. Indeed, in many non-equilibrium theories, such as Rational Extended Thermodynamics (RET) \cite{RET,Beyond,Newbook}, system \eqref{conservation} contains $M$ conservation laws. Accordingly, we assume that
\begin{equation}\label{production}
\mathbf{f}(\mathbf{u}) \equiv 
\begin{pmatrix}
\mathbf{0} \\
\mathbf{g}(\mathbf{u})
\end{pmatrix},
\qquad 
\mathbf{g} \in \mathbb{R}^{N-M}.
\end{equation}

The coupling condition introduced by Shizuta and Kawashima (the K-condition) ensures that the dissipation acting on the second block also affects the first block of equations. This condition plays a fundamental role in the global existence of smooth solutions and reads as follows.

\begin{definition}[K-condition]
On the equilibrium manifold, any characteristic right eigenvector $\mathbf{d}^{(i)}$ of the system \eqref{conservation}, \eqref{production} does not belong to the kernel of $\nabla \mathbf{f}$ evaluated at equilibrium, namely
\begin{equation}  \label{kawashima}
\nabla \mathbf{f}(\mathbf{u}_E)\cdot \mathbf{d}^{(i)} \neq 0,
\qquad \forall\, i = 1,\ldots,N,
\end{equation}
where $\mathbf{d}^{(i)}$ are defined by
\begin{equation*}
\big(\mathbf{A} - \lambda^{(i)} \mathbf{I}\big)\mathbf{d}^{(i)} = 0,
\qquad 
\mathbf{A} = \nabla \mathbf{G},
\end{equation*}
and $\mathbf{u}_E$ denotes an equilibrium state, i.e.
\begin{equation*}
\mathbf{f}(\mathbf{u}_E) = \mathbf{g}(\mathbf{u}_E) = 0.
\end{equation*}
\end{definition}
Here $\nabla$ denotes the gradient with respect to $\mathbf{u}$.

Assume that system \eqref{conservation} admits an entropy balance law
\begin{equation*}
\partial_t h(\mathbf{u}) + \partial_x k(\mathbf{u}) = \Sigma(\mathbf{u}) \leq 0,
\end{equation*}
with $h(\mathbf{u})$ strictly convex. If, in addition, the system is strictly dissipative (see \cite{arma,shock} for precise definitions), then the K-condition is a sufficient condition for the global existence of smooth solutions for small initial data. In one space dimension this was first proved by Hanouzet and Natalini \cite{nat}, and in several space dimensions by Yong \cite{wen}. Their result can be summarized as follows.

\begin{theorem}[Global existence]
Assume that system \eqref{conservation} is strictly dissipative and that the K-condition holds. Then there exists $\delta>0$ such that, if
\[
\|\mathbf{u}(x,0)\|_{2} \leq \delta,
\]
there exists a unique global smooth solution satisfying
\[
\mathbf{u} \in \mathcal{C}^{0}\!\left([0,\infty); H^{2}(\mathbb{R})\right)
\cap
\mathcal{C}^{1}\!\left([0,\infty); H^{1}(\mathbb{R})\right).
\]
\end{theorem}

Moreover, in the one-dimensional case, Ruggeri and Serre \cite{RugSerre} proved the stability of constant equilibrium states:
\begin{theorem}[Stability of constant state]
Under the assumptions of strictly convex entropy, strict dissipativeness, genuine coupling, and zero total mass of the perturbation of the equilibrium variables, the constant equilibrium solution satisfies
\[
\|\mathbf{u}(t)\|_{2} = O\!\left(t^{-1/2}\right).
\]
\end{theorem}

In both theorems, a crucial role is played by the possibility of rewriting system \eqref{conservation} in symmetric form by introducing the \emph{main field} $\mathbf{u}' = \nabla h$, first proposed by Boillat \cite{Boil} in the classical framework and by Ruggeri and Strumia \cite{RS} in a covariant formulation.

Lou and Ruggeri \cite{Palermo} introduced a weaker form of the K-condition, which requires that \eqref{kawashima} hold only for right eigenvectors associated with genuinely nonlinear characteristic fields, that is, for those indices $i \in \{1,\ldots,N\}$ such that
\[
\nabla \lambda^{(i)} \cdot \mathbf{d}^{(i)} \big|_{E} \neq 0.
\]
In particular, $\nabla \mathbf{f}(\mathbf{u}_E)\cdot \mathbf{d}^{(i)}$ is allowed to vanish for linearly degenerate fields, as in the example of Zeng \cite{zeng}.

\section{Weak-discontinuity waves}
For a generic quasilinear hyperbolic system, it is possible to consider a particular class of solutions corresponding to the so-called
\emph{weak discontinuity waves} or, in the language of continuum mechanics, \emph{acceleration waves}.
These are moving surfaces (wave fronts) that separate space into two subdomains. Ahead of and behind the wave front, the solutions are continuous across the surface, but the field exhibits a discontinuity in its normal derivative.

The theory of weak discontinuity (acceleration) waves is classical and will not be recalled here; we refer to Boillat for the three-dimensional case \cite{Boillat} and to \cite{Ruggeri,Newbook} for full details in one space dimension.

In particular, we recall that along a characteristic associated with an eigenvalue $\lambda$, the jump of the normal derivative is proportional to the corresponding right eigenvector $\mathbf d$,
\begin{equation}\label{Saltino}
    \bm{\Pi}=\llbracket \mathbf{u}_x \rrbracket
    = \Pi\, \mathbf d(\mathbf u_0),
\end{equation}
where $\mathbf u_0$ is the unperturbed state in which the wave propagates and the double brackets denote the jump across the wave front.

The scalar amplitude $\Pi$ satisfies a Bernoulli equation of the form
\begin{equation}
\frac{d\Pi}{dt}+a(t)\Pi^{2}+b(t)\Pi=0 ,
\label{bern}
\end{equation}
where $d/dt$ denotes the derivative with respect to time along the characteristic under consideration.
The qualitative properties of \eqref{bern} were analyzed in \cite{Ruggeri,Newbook}.

We restrict here to the case in which the unperturbed state is a constant equilibrium
\[
\mathbf u_0=\mathbf u_E ,\qquad \mathbf f(\mathbf u_E)=0 .
\]
In this case  the coefficients in \eqref{bern} are constant and one finds \cite{Ruggeri}
\begin{equation}\label{aabb}
a=(\nabla\lambda\cdot\mathbf d)_E,\qquad 
b=-(\mathbf l\cdot\nabla\mathbf f\cdot\mathbf d)_E ,
%\label{stabc_red}
\end{equation}
where $\mathbf d$ and $\mathbf l$ are respectively the right and left eigenvectors of $ \mathbf A$ associated with $\lambda$, normalized by $\mathbf l\cdot\mathbf d=1$.

For genuinely nonlinear waves ($a\neq0$), the solution of \eqref{bern} may blow up in finite time unless the coefficient $b$ is positive. In this case, the zero solution $\Pi=0$ is asymptotically stable if and only if \cite{Ruggeri}
\begin{equation*}
b=-(\mathbf l\cdot\nabla\mathbf f\cdot\mathbf d)_E>0 ,
%\label{stabb_red}
\end{equation*}
which coincides with what Dafermos calls weak dissipation along each characteristic field \cite{Daf1,Daf2}.

The solution of \eqref{bern} is explicitly given by
\begin{equation*}
\Pi(t)=\frac{\Pi_0 e^{-b t}}{1+\dfrac{a}{b}\Pi_0\left(1-e^{-b t}\right)} .
\end{equation*}
Therefore, assuming for example $a<0$, if the initial data satisfy
\begin{equation}\label{critical}
\Pi_0<0, \quad \text{or} \quad  0<\Pi_0<\Pi_{\rm cr},\quad 
\Pi_{\rm cr}=\frac{b}{|a|},
\end{equation}
the solution exists globally in time and converges to zero.  
Conversely, the solution blows up at the critical time
\begin{equation*}
t_c=-\frac{1}{b}\ln\!\left(1-\frac{\Pi_{\rm cr}}{ \Pi_0}\right), \quad \text{if} 
\quad \Pi_0> \Pi_{\rm cr}>0.
\end{equation*}
If instead $b=0$, the solution becomes
\begin{equation*}
    \Pi(t) = \frac{\Pi_0}{1 + a \,\Pi_0\, t},
\end{equation*}
and a critical time always occurs if $\Pi_0>0$.

Hence, for genuinely nonlinear characteristic fields, the K--condition is a necessary condition for the persistence in time of weak discontinuity waves with sufficiently small initial amplitude.

If instead the field is linearly degenerate ($a=0$), equation \eqref{bern} becomes linear and no critical time occurs even when $b=0$; therefore, in this case the K--condition is not necessary.

Introducing the operator $\delta=\jump{\partial/\partial x}$, for which $\delta\mathbf u \propto \mathbf d$, the  K--condition can be written as
\begin{equation}\label{Kweak}
   \delta\mathbf f\big|_E=(\nabla\mathbf f \cdot \delta\mathbf u)_E
\propto (\nabla\mathbf f \cdot \mathbf d)_E \neq 0. 
\end{equation}
If this condition holds for all characteristic fields, including the linearly degenerate ones, then for sufficiently smooth initial data the existence of a global solution follows, since the condition is sufficient.  
If instead, according to the weaker K--condition introduced by Lou and Ruggeri \cite{Palermo}, condition \eqref{Kweak} holds only for genuinely nonlinear fields, then it is a necessary but not sufficient condition for the existence of global solutions.

The physical meaning of \eqref{Kweak} is that, even if the production term vanishes at equilibrium, $\mathbf f(\mathbf u_E)=0$, genuinely nonlinear discontinuity waves transport a nonvanishing jump of the normal derivative of the production term.

\begin{remark}\label{remarkino}
Let $\mathbf w$ denote the non--equilibrium variables and assume that
$\mathbf f(\mathbf w):\mathbb{R}^N\to\mathbb{R}^N$ is differentiable at $\mathbf 0$ and satisfies
\[
\mathbf f(\mathbf 0)=\mathbf 0 .
\]
Then
\[
\nabla\mathbf f(\mathbf 0)\neq 0
\]
if and only if there exists a nonzero linear map $\mathbf B:\mathbb{R}^N\to\mathbb{R}^N$ such that
\[
\mathbf f(\mathbf w)=\mathbf B\,\mathbf w+o(|\mathbf w|)\qquad \text{as } \mathbf w\to 0 .
\]
In other words, the K-condition (and also the weaker version) requires that $\mathbf f(\mathbf w)$ possess a non--vanishing linear part near equilibrium.
\end{remark}

\begin{remark}[Singular limit for $b$]
Until now we have assumed $b$ to be a finite  constant. In the non--Newtonian case discussed later, the coefficient $b$ may become arbitrarily large.  
Suppose that $b(\epsilon)$ has a singular power--law dependence on a small constant  parameter $(\epsilon>0)$:
\[
b(\epsilon) = \frac{b_0}{\epsilon^n}, \qquad b_0>0, \quad n>1.
\]
The critical amplitude \eqref{critical} becomes
\[
\Pi_{\rm cr} = \frac{b(\epsilon)}{|a|} = \frac{b_0}{|a| \, \epsilon^n}. 
\]
In the singular limit $\epsilon \to 0$ one finds
$
\Pi_{\rm cr} \to +\infty, \, t_c \to +\infty,
$
so that for any finite initial amplitude $\Pi_0$ the solution exists globally in time. Moreover, the solution decays exponentially on a fast time scale:
\[
\Pi(t) \sim \Pi_0 \, \exp\!\Big(- b(\epsilon) t\Big)
= \Pi_0 \, \exp\!\Big(- \frac{b_0}{\epsilon^n} t \Big), \quad t>0,
\]
and the dominant decay time scale is 
\[
\tau_{\rm decay} \sim \frac{\epsilon^n}{b_0} \to 0
\qquad \text{as } \epsilon \to 0.
\]

Hence, in this singular limit, weak discontinuity waves are strongly damped, and for $\epsilon \to 0$ the initial discontinuity is instantaneously regularized for $t>0$.
This provides a rigorous framework for subsequent applications to non--Newtonian fluids, where $b$ depends on the internal variable $\epsilon$ in a similar power--law fashion.
\end{remark}

\section{Nonlinear model for uniaxial isothermal viscoelasticity and non-Newtonian fluids}

In this section we summarize the mathematical model, based on the principles of Rational Extended Thermodynamics (RET), used to describe a viscoelastic solid and a non-Newtonian fluid.

Recently, Ruggeri~\cite{RuggeriV} revisited earlier formulations within Rational Extended Thermodynamics and proposed a simple nonlinear one-dimensional model for isothermal viscoelasticity.  
The model is characterized by the introduction of a nonequilibrium field, treated as an independent variable associated with viscous effects, which satisfies a balance law coupled with the momentum equation.  
The resulting closed system is uniquely determined by the entropy principle.  
This guarantees thermodynamic consistency and, under the usual convexity assumptions on the entropy, the hyperbolicity of the governing equations.  
Within an isothermal framework, this formulation captures essential features of nonlinear viscoelastic response while preserving a clear and robust mathematical structure.  
Stress relaxation was studied in~\cite{AAR}, while applications to shock structure waves were investigated in~\cite{ART}.

More recently, the same theoretical framework has been shown to encompass a broader class of materials.  
In particular, in~\cite{Ruggeri2025} it was demonstrated that, by modifying the production term while keeping the overall structure of the model unchanged, the theory can be naturally interpreted as a model for non-Newtonian fluids with a finite relaxation time.  
This result highlights a close connection between viscoelasticity and non-Newtonian fluid behavior and suggests that both can be viewed as different manifestations of the same underlying thermodynamic mechanism.  
An extension to the non-isothermal case was recently considered by Arima and Ruggeri~\cite{AR-Raja}.

The system of balance laws in Lagrangian variables derived from the universal principles of RET~\cite{RuggeriV} reads:
\begin{align}
\label{balance}
\begin{split}
 & \rho^* \frac{\partial v}{\partial t} - \frac{\partial}{\partial X}\big(T(F) + \sigma\big) = 0, \\
 & \frac{\partial F}{\partial t} - \frac{\partial v}{\partial X} = 0, \\
 & \frac{\partial Z(\sigma)}{\partial t} - \frac{\partial v}{\partial X} = P(F,\sigma).
\end{split}
\end{align}
Here, $X$ denotes the position in the reference configuration, $\rho^*$ the mass density, $v$ the velocity, and $F$ the deformation gradient. 
Moreover, $T$ and $\sigma$ represent the elastic and viscous stresses, respectively, both defined as components of the first Piola--Kirchhoff stress tensor, and $P$ is the production term, whose explicit form will be specified later.

Under the assumption that the internal energy can be decomposed into the sum of an elastic part $\eE(F)$ and a viscous part $\eV(\sigma)$, any smooth solution of~\eqref{balance} also satisfies the supplementary energy balance law
\begin{equation}\label{energia}
\rho^* \frac{\partial}{\partial t}\!\left(\frac{v^2}{2} + \eE(F) + \eV(\sigma)\right)
- \frac{\partial}{\partial X}\!\left\{\big(T(F) + \sigma\big)v\right\}
= {\cal E} \leq 0,
\end{equation}
provided that $T(F)$ and the function $Z(\sigma)$ are chosen as 
\begin{align}\label{Zzz}
\begin{split}
    & T(F) = {\cal W}'(F), \qquad Z(\sigma) = \int \omega(\sigma)\, d\sigma, \\
& \text{with} \quad {\cal W}(F) = \rho^* \eE(F), \qquad
\omega(\sigma) = \rho^* \frac{{\eV}'(\sigma)}{\sigma},
\end{split}
\end{align} 
and the production term satisfies
\begin{equation}\label{reduced}
{\cal E} = \sigma\, P(F,\sigma)\leq 0, \qquad \forall\, (F,\sigma).
\end{equation}
When $\sigma = 0$, the system is in an equilibrium state and the energy production attains its maximum value ${\cal E}=0$.
Therefore, we must require
\begin{equation}\label{PPF}
P(F,0)=0, \qquad P_F(F,0)=0, \qquad P_\sigma(F,0) \leq 0 .
\end{equation}
Here and in the following, the prime denotes differentiation with respect to the corresponding independent variable, while partial derivatives with respect to different variables are denoted by subscripts.

The explicit form of the production term $P$ depends on whether the material is a viscoelastic solid or a non-Newtonian fluid and will be discussed in the next sections.

We stress that inequality~\eqref{energia} expresses the \emph{entropy principle} in the isothermal setting.

The system is closed once the constitutive functions $\eE(F)$ and $\eV(\sigma)$, together with the production term $P(F,\sigma)$, are prescribed.

We observe that the first two equations in~\eqref{balance} represent the momentum balance written in first-order form, while the third equation is a balance law for the viscous stress, in accordance with the guiding principle of RET that each new flux variable is associated with its own balance equation.  
Within this framework, the evolution equation for $\sigma$ should not be interpreted as a constitutive relation but rather as a balance law, like the others, and therefore does not need to satisfy the principle of material frame indifference, which must be imposed only on the true constitutive equations.
This viewpoint is consistent with the conjecture in~\cite{Ruggeri_Can} that physical phenomena are fundamentally governed by hyperbolic balance laws endowed with constitutive equations of local type.

\section{Acceleration waves for the present model}\label{sect5}
The system can be written in the form \eqref{conservation} with the identification  
(the superscript $\top$ denotes transpose and the space variable is now $X$)
\begin{equation*}
	\mathbf{u}\equiv (\rstar v,F,Z(\sigma))^\top, \quad 
	\mathbf{G} \equiv \left(-\bigl(T(F)+\sigma\bigr),-v,-v\right)^\top,
	\quad 
	\mathbf{f}\equiv \left(0,0, P(F,\sigma)\right)^\top.
\end{equation*}

As for acceleration waves, the balance form is not needed. It is therefore convenient to rewrite system \eqref{conservation} in the equivalent quasi-linear form
\begin{equation}\label{quasil}
	\partial_t \mathbf{u} + \mathbf{A}(\mathbf{u})\,\partial_X\mathbf{u} = \mathbf{f}(\mathbf{u}),
\end{equation}
by substituting the field and production terms as
\begin{equation}\label{fieldd}
	\mathbf{u} \equiv (v,F,\sigma)^\top, \qquad 
	\mathbf{f} \equiv \left(0,0,\frac{P(F,\sigma)}{\omega(\sigma)}\right)^\top.
\end{equation}
With this choice of the field \eqref{fieldd}$_1$, the characteristic matrix \eqref{quasil} associated with the system \eqref{balance} becomes
\[
\mathbf{A}\equiv  
\left(
\begin{array}{ccc}
	0 & -\dfrac{{\cal W}''(F)}{\rstar} & -\dfrac{1}{\rstar} \\
	-1 & 0 & 0 \\
	-\dfrac{1}{\omega(\sigma)} & 0 & 0
\end{array}
\right).
\]
The eigenvalues of $\mathbf{A}$ are $0,\pm \lambda$, where
\begin{equation*}
  \lambda=  \frac{\sqrt{\omega (\sigma ) {\cal{W}}^{''}(F)+1}}{\sqrt{\rstar}\sqrt{\omega(\sigma)}}. 
\end{equation*}
The corresponding right and left eigenvectors of $\mathbf{A}$ associated with the maximum speed $\lambda$ are
\begin{equation}\label{del}
\mathbf{d}\equiv \frac{1}{\lambda }\left(-1,\frac{1}{\lambda},\frac{1}{\lambda\, \omega (\sigma)}\right)^\top, 
\quad 
\mathbf{l}\equiv \frac{1}{2}\left(\lambda, \frac{{\cal{W}}^{''}(F)}{\rstar},-\frac{1}{\rstar}\right),
\end{equation}
chosen so that $\mathbf{l}\cdot\mathbf{d}=1$.

Moreover, with this choice of $\mathbf{d}$, relation \eqref{Saltino} yields
\begin{equation*}
(\jump{v_X},\jump{F_X},\jump{\sigma_X})^\top=\Pi\,\mathbf{d}_0,
\end{equation*}
so that, from \eqref{del}, in particular $\jump{v_X}=-\Pi/\lambda_0$.
Let ${\cal{G}}$ denote the jump of acceleration:
\begin{equation*}
    {\cal{G}}=\jump{v_t}= -\lambda_0 \jump{v_X}= \Pi.
\end{equation*}
Therefore, with this normalization of the right eigenvector, the amplitude $\Pi$ coincides with the acceleration jump ${\cal{G}}$.

Straightforward computations give
\[
\bm{\nabla} \lambda = \frac{1}{2 \lambda \rstar} 
\left(0,{\cal{W}}^{'''}(F),-\frac{\omega '(\sigma )}{\omega (\sigma )^2}\right),
\]
and
\[
\bm{\nabla} \mathbf{f} =
\left(
\begin{array}{ccc}
 0 & 0 & 0 \\
 0 & 0 & 0 \\
 0 & \dfrac{P_F(F,\sigma)}{\omega (\sigma)} &
     \dfrac{P_\sigma(F,\sigma)\omega (\sigma)-P(F,\sigma)\omega'(\sigma)}{\omega (\sigma)^2}
\end{array}
\right).
\]

Assuming that the wave propagates in an undeformed equilibrium state, $F=1$ and $\sigma=0$, and using \eqref{PPF} and  \eqref{aabb} we obtain
\begin{equation}\label{abab}
 a=   \frac{-\omega '(0)+\omega (0)^3 {\cal{W}}^{'''}(1)}{2
   \lambda_0 ^3 \rstar \omega (0)^3}, 
\qquad 
 b=-\frac{P_\sigma(1,0)}{2 \rstar \lambda_0 ^2
   \omega (0 )^2},
\end{equation}
and
\begin{equation}\label{Gcritical}
  \lambda_0=  \frac{\sqrt{\omega (0 )
   {\cal{W}}^{''}(1)+1}}{\sqrt{\rstar}\sqrt{\omega(0)}}, 
\qquad 
 {\cal{G}}_{\rm cr}=\frac{\lambda_0 \omega (0)
   P_\sigma(1,0)}{\omega (0)^3
   {\cal{W}}^{'''}(1)-\omega '(0)}.
\end{equation}

In agreement with the physical examples discussed below, we assume $\omega'(0)\geq 0$ and ${\cal{W}}^{'''}(1)<0$, which implies $a<0$.
We distinguish three relevant cases.

\medskip
\noindent\textbf{Case 1: $P_\sigma(1,0)<0$ (hence $b>0$ and finite).}  
If the initial acceleration jump satisfies $|{\cal{G}}_0|<{\cal{G}}_{\rm cr}$, then a global solution exists and the K--condition holds.  
Finite-time blow-up occurs if and only if
\begin{equation*}
{\cal{G}}_0 > {\cal{G}}_{\rm cr} > 0,
\end{equation*}
with ${\cal{G}}_{\rm cr}$ given by \eqref{Gcritical}$_2$, and the critical time
\begin{equation}\label{tccase1}
t_c=-\frac{1}{b}\ln\!\left(1-\frac{{\cal{G}}_{\rm cr}}{ {\cal{G}}_0}\right). 
\end{equation}

\medskip
\noindent\textbf{Case 2: $P_\sigma(1,0)=0$ (hence $b=0$, degenerate case).}  
The K--condition is violated and, since $a<0$, any positive initial amplitude ${\cal{G}}_0>0$ leads to finite-time blow-up at
\begin{equation*}
t_c = -\frac{1}{a {\cal{G}}_0} > 0.
\end{equation*}

\medskip
\noindent\textbf{Case 3: $P_\sigma(1,0)\sim -\varepsilon^{-n}$, $n>1$, where $\varepsilon$ is a 
small parameter (thus $b(\varepsilon)\to + \infty$ as $\varepsilon\to 0$).}  

In this singular limit, the quadratic term is negligible at early times and the evolution reduces to
\begin{equation*}
\frac{d{\cal{G}}}{dt} + b(\varepsilon) {\cal{G}} \approx 0,
\qquad 
{\cal{G}}(t) \approx {\cal{G}}_0 e^{-b(\varepsilon) t}.
\end{equation*}

\noindent\textbf{Observations:}
\begin{itemize}
    \item The critical amplitude ${\cal{G}}_{\rm cr}=b(\varepsilon)/|a|\to\infty$ as $\varepsilon\to 0$.
    \item For any fixed ${\cal{G}}_0$, the solution remains bounded and decays rapidly.
    \item Finite-time blow-up disappears in the singular limit: linear damping dominates and the system becomes effectively stable.
\end{itemize}

\medskip
\noindent\textbf{Summary:}
\begin{itemize}
    \item For finite positive $b(0)$, blow-up occurs above the threshold ${\cal{G}}_{\rm cr}$.
    \item For $b(0)=0$, all positive initial amplitudes lead to blow-up.
    \item For very large $b(\varepsilon)$ (singular limit), blow-up is suppressed and solutions decay rapidly. In the limit $\varepsilon
    \to 0$, the initial discontinuity is instantaneously regularized for $t>0$.
\end{itemize}

\section{Acceleration waves  in viscoelastic solids under uniaxial deformation}
In this section, we specialize the previous general results to the model of a viscoelastic solid.
For a uniaxial deformation, we set
\begin{equation}\label{Funi}
    \mathbf{F} = \mathrm{diag}(F, F_\perp, F_\perp),
\end{equation}
and impose the zero transverse stress condition $T_{22} = T_{33} = 0$.  
Near the undeformed configuration, the lateral stretch satisfies
\begin{equation}\label{Fperp}
    F_\perp = F^{-\bar{\nu}_0}
\simeq 1 - \bar{\nu}_0 (F - 1) + O\!\left((F - 1)^2\right),
\end{equation}
where $\bar{\nu}_0$ denotes the Poisson ratio.  

In the present case \cite{AAR,ART}, the longitudinal stress is given by
\[
P(F,\sigma) = -\frac{\sigma}{\mu(F)}, 
\qquad 
\mu(F) = \frac{\mu_0}{F^{1 + 2 \bar{\nu}_0}}, 
\qquad  \mu_0 > 0.
\]
It was proved in \cite{RuggeriV} that, for quadratic potentials
\begin{equation}\label{elin}
{\cal{W}}(F)=\rho^* \eE = \frac{E_1}{2}(F-1)^2,
\qquad
\rho^* \eV(\sigma)= \frac{\sigma^2}{2 E_2},
\end{equation}
the balance equation \eqref{balance} for the total Piola stress 
$S=T+\sigma$ reduces to the Zener (standard linear solid) model \cite{Amabili}, namely
\begin{equation}\label{zen}
 \frac{1}{E_2} \frac{\partial S}{\partial t}
 -\left(1+\frac{E_1}{E_2}\right)\frac{\partial \varepsilon}{\partial t}
 = \frac{-S +E_1 \varepsilon}{\mu_0}, 
 \qquad \mu_0=\mu(1),
\end{equation}
where $\varepsilon=F-1$ is the linear Green--Lagrange strain.  
Here $E_1$ is the elastic modulus of the spring in parallel (long-term elastic response), and $E_2$ is the elastic modulus of the spring in series with the dashpot (short-term elastic response).
From \eqref{elin} we obtain
\[
\omega(0)=\frac{1}{E_2}, 
\qquad 
{\cal{W}}^{''}(1)=E_1, 
\qquad 
P_\sigma(1,0)= -\frac{1}{\mu_0}.
\]
However, in order to compute explicitly $a$ and hence ${\cal{G}}_{\rm cr}$, we also need $\omega'(0)$ and ${\cal{W}}^{'''}(1)$, which requires a third-order approximation of the potentials.

Concerning the viscous energy $\eV(\sigma)$, although nonlinear models have been proposed in specific cases \cite{Fractional,AAR}, they do not admit a simple closed form.  
For this reason, in the present analysis we adopt the quadratic viscous energy \eqref{elin}$_2$.

For the elastic potential, the precise expression depends on the chosen constitutive model. Nevertheless, since ${\cal{W}}$ and all its derivatives have the same physical dimension as $E_1$, we may write, near equilibrium,
\begin{equation}\label{Wterzo}
    {\cal{W}}(F)=\frac12 E_1 (F-1)^2 
- \frac13 E_1 R (F-1)^3 
+ O\!\left((F-1)^4\right),
\end{equation}
where the dimensionless constant $R$ depends on the particular form of the potential and will be shown to be positive for the models considered below.  
Hence,
\begin{equation*}
\omega'(0)=0,
\qquad 
{\cal{W}}^{'''}(1)=-2 E_1 R<0.
\end{equation*}
In this case the characteristic velocity in the unperturbed state and the coefficients $a$ and $b$ read:
\begin{equation}\label{abcase1}
\lambda_0=\sqrt{\frac{E_1+E_2}{\rstar}}, \quad a = -\frac{\sqrt{\rho^*} \,E_1 R}{(E_1 + E_2)^{3/2}}<0, 
\quad
b = \frac{E_2}{2 \tau_0 (E_1 + E_2)}>0,
\end{equation}
where the relaxation time $\tau_0$ is related to the viscosity coefficient $\mu_0$ by (see \eqref{zen})
\begin{equation}\label{ttt}
    \tau_0= \frac{\mu_0}{E_2}.
\end{equation}
Since $b>0$, we are in case~1) of Section~\ref{sect5}.  
The critical initial amplitude for finite-time blow-up is therefore
\begin{equation}\label{Gccase1}
{\cal{G}}_{\rm cr}
=\frac{b}{|a|}
= \frac{1}{2 R \tau_0}\frac{E_2}{ E_1}\sqrt{\frac{E_1 + E_2}{\rho^*}}.
\end{equation}

\noindent
For any ${\cal{G}}_0>{\cal{G}}_{\rm cr}$, the critical time at which ${\cal{G}}(t)$ blows up is given by \eqref{tccase1}, with $b$ and ${\cal{G}}_{\rm cr}$ given in \eqref{abcase1}$_3$ and \eqref{Gccase1}, respectively.

In~\cite{RuggeriV} it was shown that the contact waves $\lambda=0$ also satisfy \eqref{Kweak}, and therefore the present model satisfies the full K-condition and not only its weaker version.
\subsection{Evaluation of $R$ with the Mooney--Rivlin potential}

The only quantity depending on the particular constitutive equation adopted for the elastic response is $R$.  
To obtain an order-of-magnitude estimate of the critical acceleration , we adopt the well-known compressible Mooney--Rivlin model for the strain energy density (see, e.g., \cite{Hanz,Nrug}):
\begin{align} \label{Mooney-Rivlin}
  W =  C_1 \left(\bar{I}_1 - 3\right) + C_2 \left(\bar{I}_2 - 3\right) + \frac{k}{2}\left(J - 1\right)^2,
\end{align}
where $J=\det \mathbf{F}$, $C_1$ and $C_2$ are material constants associated with the deviatoric response, $\bar{I}_1$ and $\bar{I}_2$ are the first and second invariants of the unimodular deformation tensor $\bar{\mathbf{C}}$, and $k$ is a material parameter related to the compressibility of the material.  

Since $\bar{\mathbf{C}}= J^{-2/3} \mathbf{C}$ with $\mathbf{C}=\mathbf{F}^T\mathbf{F}$ and $J=\det \mathbf{F}=\det \sqrt{\mathbf{C}}$, we have (see, e.g., \cite{Hanz}):
\begin{equation*}
    \bar{I}_1 = J^{-\frac{2}{3}} I_1, \qquad 
    \bar{I}_2 = J^{-\frac{4}{3}} I_2,
\end{equation*}
where $I_1 = \operatorname{tr} \mathbf{C}$ and 
\[
I_2=\frac{1}{2}\left[(\operatorname{tr} \mathbf{C})^2-\operatorname{tr}(\mathbf{C}^2)\right]
\]
are the first and second principal invariants of $\mathbf{C}$.

Recalling \eqref{Funi}, the uniaxial stress $T_{11}$ is given by
\[
T_{11}(F,F_\perp)=\frac{\partial W}{\partial F}\Big|_{F_\perp},
\]
so that the stress along the uniaxial loading path is
\[
T(F)=T_{11}(F,F_\perp(F)),
\]
where $F_\perp(F)$ near the undeformed configuration is approximated by \eqref{Fperp}.  
The corresponding reduced uniaxial potential is defined by
\[
\mathcal{W}(F)=\int_1^F T(\tilde F)\,d\tilde F .
\]
The explicit expression of $\mathcal{W}(F)$ and its derivatives up to third order are reported in \cite{ART}.  
Considering, as in that paper, a \emph{vulcanized rubber}, constants compatible with \eqref{Mooney-Rivlin} were estimated in~\cite{2009_HorganMurphy,2021_PengHanLi} based on Penn's experimental data~\cite{1970_Penn}, namely
\begin{equation*}
    C_1 = 0.092~\mathrm{MPa}, \quad 
    C_2 = 0.237~\mathrm{MPa}, \quad 
    k = 2000.20~\mathrm{MPa}, \quad 
    \bar{\nu}_0 = 0.4998. 
\end{equation*}
Inserting these values into the expressions for the second and third derivatives given in \cite{ART}, and taking into account \eqref{Wterzo}, we obtain
\[
{\cal W}''(1) = E_1 = 2.12~\mathrm{MPa}, 
\qquad 
{\cal W}'''(1) = -2 E_1 R = -6.93~\mathrm{MPa},
\]
so that
\[
R = 1.63,
\]
To obtain an order-of-magnitude estimate, we choose
\[
E_2 = 3~\mathrm{MPa}, \qquad
\tau_0 = 0.1~\mathrm{s}, \qquad
\rho^* = 929~\mathrm{kg/m^3}.
\]
Then we obtain (with $g$ denoting the gravitational acceleration)
\[
\lambda_0 \approx 74.21~\mathrm{m/s}, \quad
a \approx -0.009~\mathrm{s/m}, \quad 
b \approx 2.93~\mathrm{s^{-1}}, \quad 
{\cal{G}}_{\rm cr} \approx 321.41~\mathrm{m/s^2} \approx 32.80\,g.
\]
The critical acceleration is extremely large; therefore, for any physically reasonable initial acceleration ${\cal{G}}_0$, the solution decays rapidly to zero as time increases.
\section{Acceleration waves in a non-Newtonian fluid}

In this section, we introduce the production term necessary to incorporate a non-Newtonian fluid with relaxation into the model, and we specialize the results for acceleration waves in this context.

As observed in~\cite{Ruggeri2025}, the compressible fluid case including non-Newtonian behavior, obtained as a hyperbolic extension of the power-law model, can be embedded into the same system~\eqref{balance} by taking $F_\perp = 1$, $T = -p$, and $F = \rho^*/\rho$, where $\rho$ is the current density and $p$ the pressure, together with a suitable choice of the production term.

To describe \emph{non-Newtonian} behaviour, we recall the power-law model
\begin{equation}\label{PowerLaw}
\boldsymbol{\sigma} = k\,\dot{\gamma}^{\,m-1}\mathbf{D}, 
\qquad 
\dot{\gamma} = 2\lVert \mathbf{D}\rVert, 
\qquad 
\lVert \mathbf{D}\rVert=\sqrt{\mathbf{D}:\mathbf{D}}
%|\mathbf{D}|=\sqrt{\mathbf{D}:\mathbf{D}},
\end{equation}
where $k>0$ is the consistency coefficient and $m>0$ is the flow index.
In one-dimensional flow, we assume that this relation also holds for compressible fluids. The inverse of~\eqref{PowerLaw} in one dimension reduces to
\begin{equation}
	v_x = \frac{v_X}{F} 
	= 2^{\frac{1}{m}-1} k^{-1/m} |\sigma|^{\frac{1-m}{m}} \sigma .
	\label{powerlaw1D}
\end{equation}

From the last equation in~\eqref{balance}, we observe that in the limit $Z_\sigma \to 0$ one has $v_X = -P$, where $P$ is arbitrary except for the residual inequality~\eqref{reduced}. 
Therefore, by comparison with \eqref{powerlaw1D}, if we choose the production term
\begin{equation}\label{produzione}
    P = -F\, 2^{\frac{1}{m}-1} k^{-1/m} |\sigma|^{\frac{1-m}{m}} \sigma ,
\end{equation}
the RET system~\eqref{balance} reproduces the power-law relation in the parabolic limit.

The resulting one-dimensional isothermal RET model reads in Lagrangian variables \cite{Ruggeri2025}:
\begin{align}
	\begin{split}
		& \rho^* v_t + (p(F) - \sigma)_X = 0, \\
		& F_t - v_X = 0, \\
		& Z_t(\sigma) - v_X 
		= -F\, 2^{\frac{1}{m}-1} k^{-1/m} |\sigma|^{\frac{1-m}{m}} \sigma .
	\end{split}
	\label{RET-powerlaw}
\end{align}
with $p(F) = -{\cal{W}}'(F)$ and $Z(\sigma)$ given in~\eqref{Zzz}.  
System~\eqref{RET-powerlaw} satisfies the energy dissipation inequality~\eqref{energia}, with
\[
	\mathcal{E} = -F\, 2^{\frac{1}{m}-1} k^{-1/m} |\sigma|^{\frac{1+m}{m}} \le 0 .
\]

We now consider, also in this case, a quadratic viscous energy~\eqref{elin}$_2$ with \eqref{ttt}, for which $\omega$ is constant and given by
\begin{equation}\label{omegaf}
    \omega= \frac{\tau_0}{\mu_0}.
\end{equation}
Moreover, we choose the pressure of an isothermal ideal gas (with ${\cal{R}}$ constant),
\[
p = {\cal{R}} \rho = {\cal{R}} \frac{\rho^*}{F}.
\]
Then
\begin{equation}\label{W2W3}
   {\cal{W}}''(1) = {\cal{R}} \rho^*, 
   \qquad 
   {\cal{W}}'''(1) = -2 {\cal{R}} \rho^* .
\end{equation}

Introducing the dimensionless number
\begin{equation}\label{mutilde}
    \tilde{\mu}_0 = 1 + \frac{\mu_0}{{\cal{R}} \rho^* \tau_0} > 1 ,
\end{equation}
from~\eqref{abab}, \eqref{Gcritical}, \eqref{omegaf}, and~\eqref{W2W3}, we obtain
\[
\lambda_0=\sqrt{{\cal{R}} \tilde{\mu}_0}, 
\qquad 
a= -\frac{1}{\sqrt{{\cal{R}}\, \tilde{\mu}_0^3}} .
\]

To evaluate $b$, we consider $\sigma \ge 0$. From~\eqref{produzione} we obtain
\begin{equation}\label{Pss}
    P_\sigma(F,\sigma) 
    = -\frac{F}{m}\, 2^{\frac{1}{m}-1} k^{-1/m} 
    \sigma^{\frac{1}{m}-1}.
\end{equation}
We now distinguish three cases.

\medskip
\noindent\textbf{Case 1: Newtonian fluid ($m=1$).}  
In this case, from~\eqref{Pss}, $P_\sigma(1,0)=-1/\mu_0$ with $k=\mu_0$. Then, from~\eqref{abab} and~\eqref{Gcritical}, we have
\[
b = \frac{\tilde{\mu}_0-1}{2 \tilde{\mu}_0 \tau_0} > 0, 
\qquad 
{\cal G}_{\rm cr} = \frac{\sqrt{{\cal{R}} \tilde{\mu}_0}}{2 \tau_0} (\tilde{\mu}_0-1) > 0.
\]
This corresponds to Case~1 of Section~\ref{sect5}. The K-condition holds, and the solution ${\cal G}(t)$ decays globally if ${\cal G}_0 < {\cal G}_{\rm cr}$. Conversely, if ${\cal G}_0 > {\cal G}_{\rm cr}$, blow-up occurs at the critical time~\eqref{tccase1}. 

The ratio
\[
\frac{\mu_0}{{\cal{R}} \rho^* \tau_0}
\]
appearing in~\eqref{mutilde} is a dimensionless viscous--compressibility parameter measuring the ratio of viscous to elastic stresses over the relaxation time~$\tau_0$. For Newtonian fluids this ratio is typically very small, so that viscous effects represent only a small perturbation to sound propagation~\cite{Landau,Whitham}.  
Therefore, in~\eqref{mutilde} one has $\tilde{\mu}_0 \simeq 1$, and consequently $b$ and ${\cal G}_{\rm cr}$ are very small. Hence, for finite initial acceleration jumps, the solution grows and blows up at a finite critical time.

\medskip
\noindent\textbf{Case 2: Shear-thinning fluid ($m<1$).}  
Taking into account Remark~\ref{remarkino} and equation~\eqref{Pss}, we have $P_\sigma(1,0)=0$, which implies that we are in Case~2 of Section~\ref{sect5}. Then $b=0$, the K-condition is violated, and for any ${\cal G}_0>0$ a critical time exists:
\[
t_c = \frac{\sqrt{{\cal{R}} \, \tilde{\mu}_0^3}}{{\cal G}_0}.
\]

\medskip
\noindent\textbf{Case 3: Shear-thickening fluid ($m>1$).}   
From~\eqref{Pss}, we have
\[
\lim_{\sigma \to 0} P_\sigma(1,\sigma) = -\infty.
\]
To eliminate this singularity, we introduce a regularized production term, as commonly done in power-law models, by means of a small parameter $\varepsilon$.  
For $m > 1$, we replace the expression for $P$ given in~\eqref{produzione} by
\begin{equation*}
    P = -F \, 2^{\frac{1}{m}-1} k^{-1/m} |\varepsilon + \sigma|^{\frac{1-m}{m}} \sigma,
\end{equation*}
and obtain
\[
P_\sigma(1,0) 
= -\frac{1}{2^n k^{1/m} \varepsilon^n}, 
\qquad 
n = \frac{m-1}{m} > 0.
\]
In this case, we are in Case~3 of Section~\ref{sect5}. The critical amplitude ${\cal G}_{\rm cr} \to \infty$ and $b \to \infty$ as $\varepsilon \to 0$. Therefore, for any fixed ${\cal G}_0$, the solution remains bounded and decays rapidly.  
In the limit $\varepsilon \to 0$, the initial acceleration jump is instantaneously regularized.

\section{Concluding Remarks}
The analysis presented in this work highlights the interplay between two competing mechanisms in hyperbolic systems with relaxation: the nonlinearity associated with the hyperbolic character, represented by the quadratic term with coefficient $a$, and the dissipative effects, described by the linear term with coefficient $b$.
When dissipation is sufficiently strong, or when nonlinearity is weak, the critical threshold becomes very large. In this regime, solutions exist for essentially all times, and an initial acceleration discontinuity, provided it is not too large, decays monotonically. Conversely, when dissipation is weak or nonlinear effects dominate, the solution exhibits finite-time blow-up even for small initial data.
This qualitative behavior is clearly illustrated by the two classes of examples considered here. For viscoelastic solids and shear-thickening non-Newtonian fluids, dissipation dominates and the amplitude of acceleration waves decreases over time. For shear-thinning non-Newtonian fluids and, in practice, Newtonian fluids, the effective viscosity is insufficient to counterbalance nonlinear effects, and the acceleration amplitude grows, leading to blow-up at a finite critical time.
These results underscore the physical significance of the K-condition and its weaker form. They also emphasize the fundamental role played by the constitutive structure of the material in determining the long-time behavior of acceleration waves, showing how dissipation and nonlinearity jointly control their evolution.

\begin{acknowledgements}
The work was carried out within the activities of the Italian National Group of Mathematical Physics (GNFM) of the Italian National Institute of Higher Mathematics (INdAM).
\end{acknowledgements}

\section*{Conflict of Interest}
The author declares that he has no conflict of interest.

\end{document}